\documentclass[showpacs,aps,prd,nofootinbib,floatfix,amsmath,amssymb]{revtex4}
\usepackage{graphicx}
\begin{document}

\makeatletter
\newbox\slashbox \setbox\slashbox=\hbox{$/$}
\newbox\Slashbox \setbox\Slashbox=\hbox{\large$/$}
\def\pFMslash#1{\setbox\@tempboxa=\hbox{$#1$}
  \@tempdima=0.5\wd\slashbox \advance\@tempdima 0.5\wd\@tempboxa
  \copy\slashbox \kern-\@tempdima \box\@tempboxa}
\def\pFMSlash#1{\setbox\@tempboxa=\hbox{$#1$}
  \@tempdima=0.5\wd\Slashbox \advance\@tempdima 0.5\wd\@tempboxa
  \copy\Slashbox \kern-\@tempdima \box\@tempboxa}
\def\FMslash{\protect\pFMslash}
\def\FMSlash{\protect\pFMSlash}
\def\miss#1{\ifmmode{/\mkern-11mu #1}\else{${/\mkern-11mu #1}$}\fi}
\makeatother

\title{Bounding the  $B_s\to \gamma \gamma$ decay  from Higgs mediated FCNC transitions}
\author{J. I. Aranda$^{(a)}$, J.  Monta\~no$^{(b)}$, F. Ram\'\i rez-Zavaleta$^{(a)}$, J. J. Toscano$^{(c)}$, and E. S. Tututi$^{(a)}$}
\address{$^{(a)}$Facultad de Ciencias F\'\i sico Matem\' aticas,
Universidad Michoacana de San Nicol\' as de
Hidalgo, Avenida Francisco J. M\' ujica S/N, 58060, Morelia, Michoac\'an, M\' exico. \\
$^{(b)}$Departamento de F\'\i sica, Universidad de Guanajuato,
Campus Leon, C.P. 37150, Le\' on, Guanajuato, M\' exico.\\
$^{(c)}$Facultad de Ciencias F\'{\i}sico Matem\'aticas,
Benem\'erita Universidad Aut\'onoma de Puebla, Apartado Postal
1152, Puebla, Puebla, M\'exico.}

\begin{abstract}
The Higgs mediated flavor violating bottom-strange quarks transitions induced at the one-loop level by a nondiagonal $Hbs$ coupling are studied within the context of an effective Yukawa sector that comprises $SU_L(2)\times U_Y(1)$-invariant operators of up to dimension-six. The most recent experimental result on $B\to X_s\gamma$ with hard photons is employed to constraint the $Hbs$ vertex, which is used to estimate the branching ratio for the $B_s\to \gamma\gamma$ decay. It is found that the $B_s\to \gamma\gamma$ decay can reach a branching ratio of the order of $4\times 10^{-8}$, which is 2 orders of magnitude more stringent than the current experimental limit.
\end{abstract}


\maketitle

\section{Introduction}
\label{Int} Radiative $B$ decays have attracted considerable attention in the last years. The rich phenomenology of weak meson decays~\cite{Review} provides an excellent laboratory to probe effects of physics beyond the Standard Model (SM). In particular, suppressed observables such as $B\to X_s\gamma$, potentially sensitive to new physics effects, has been measured with good accuracy, showing no deviations from the SM. This means that this observable can provide stringent constraints on physics beyond the electroweak scale. In fact, the rare $b\to s\gamma$ decay has shown to be very sensitive to possible new physics effects in diverse scenarios~\cite{Hewett}. It results that the leading contribution to $B\to X_s\gamma$ decay with a hard photon is dominated by $b\to s\gamma$ process. The current experimental value,  which is given by the Heavy Flavor Averaging Group (HFAG)~\cite{HFAG} along
with the BABAR, Belle and CLEO  collaborations, is $Br(B \to X_s\gamma) = (3.52\pm 0.23\pm 0.09)\times10^{-4}$ for a photon energy $E_\gamma>1.6$ GeV. On the theoretical side, the SM prediction at the next to next leading order is $Br(B\to X_s\gamma)=(3.15\pm 0.23)\times 10^{-4}$ for $E_\gamma \gtrsim 1.6$ GeV~\cite{SMP}. This high level of coincidence between experimental and theoretical results leaves a very small room for physics beyond the SM, which means that this process can lead to strong constraints on new physics effects.

In this paper, we are interested in studying the flavor violating transitions $b\to s\gamma$ and $b\to s\gamma \gamma$ mediated by a SM-like Higgs boson within the context of extended Yukawa sectors, which are always present within the SM with additional $SU_L(2)$-Higgs multiplets or in larger gauge groups. Some processes naturally associated with flavor violation  could be significantly impacted by extended Yukawa sectors, as it is expected that more
complicated Higgs sectors tend to favor this class of new physics effects. We will assume that the flavor violating decays $b\to s\gamma$ and $b\to s\gamma \gamma$ are mediated by a virtual neutral Higgs boson with a mass of the order of magnitude of the Fermi scale $v\approx 246$ GeV. However, instead of tackling the problem in a specific model, we will adopt a model independent approach by using the effective Lagrangian technique~\cite{EL}, which is an appropriate scheme to study those
processes that are suppressed or forbidden in the SM.  As it has been shown in Refs.~\cite{T1,T2,T2-1,T3,PV}, it is not necessary to introduce new degrees of freedom in order to generate flavor violation at the level of classical action, the introduction of operators of dimension higher than four will be enough. We will see below that an effective Yukawa sector that incorporates $SU_L(2)\times U_Y(1)$-invariants of up to dimension six is enough to reproduce, in a model independent manner, the main features that are common to extended Yukawa sectors, such as the presence of flavor and CP violation. Although theories beyond the SM require more complicated Higgs sectors that include new physical scalars, we stress that our approach for studying flavor violation mediated by a relatively light scalar particle is sufficiently general to incorporate the most relevant aspects of extended theories. As in most cases, it is always possible to identify in an appropriate limit a SM-like Higgs boson whose couplings to pairs of $W$ and $Z$ bosons coincide with those given in the minimal SM. This is the case of the most general version of the two-Higgs doublet model (THDM-III)~\cite{THDM-III} and  multi-Higgs models that comprise additional multiplets of $SU_L(2)\times U_Y(1)$ or scalar representations of larger gauge groups. Our approach also cover more exotic formulations of flavor violation, as the so-called familons models~\cite{Familons} or theories that involves an Abelian flavor symmetry~\cite{AFS}. In this way, our results will be applicable to a wide variety of models that predict scalar-mediated flavor changing neutral currents (FCNC). Besides its model independence, our framework has the advantage that it involves an
equal or even less number of unknown parameters than those usually appearing in specific extended Yukawa sectors.

Our main goal in this work is to use the experimental data on the $B\to X_s\gamma$ decay to constraint the flavor violating $Hbs$ vertex induced by the effective Yukawa sector described above and to establish the amplitude for the $b\to s\gamma \gamma$ decay within this context. Then we will use these results to predict the branching ratio for the $B_s\to \gamma \gamma$ transition. The amplitude for the $b\to s\gamma \gamma$ transition has already been calculated within the context of the SM and used to estimate the branching ratio for the $B_s\to \gamma \gamma$ decay, which, without QCD corrections, was found to be of the order of $10^{-7}$~\cite{Yao, SQA}. Subsequent studies showed that the next order QCD corrections increase the branching ratio up to a value of $\sim (1.0-1-2)\times 10^{-6}$~\cite{NOQCD}. Long-distance effects, where the two photons are emitted from intermediate states, such as $B_s\to \phi \gamma \to \gamma \gamma$~\cite{LD1}, $B_s\to \Psi \phi \to \gamma \gamma $~\cite{LD2} or $B_s\to D^{(*)+} D^{(*)-}\to \gamma \gamma$~\cite{LD3}, have also been considered. It was found that these processes lead to corrections of $20\%$, at the best. In contrast with this behavior, we will see that in our case the short-distance effects are marginal and that a long-distance effect mediated by the Higgs boson $H$ dominates. In this effect, the two photons are emitted from a virtual Higgs boson through the $B_s\to H^*\to \gamma \gamma$ process, in which the SM one-loop vertex $H^\ast\gamma \gamma$ play a crucial role. Beyond the SM, the $B_s\to \gamma \gamma$ decay has been studied in softly broken supersymmetry~\cite{SBSUSY}, in the two Higgs doublet model~\cite{THDM}, in presence of a four generation~\cite{FG}, and in supersymmetry with broken $R$-parity~\cite{BRPSUSY}. In general terms, the branching ratio for the $B_s\to \gamma \gamma$ decay calculated in the SM and other of its extensions is located in the range $(0.4-1.0)\times 10^{-6}$, which is one order of magnitude lower than the upper limit $Br(B_s\to \gamma \gamma)<8.7\times 10^{-6}$ obtained by the Belle experiment~\cite{BelleExp}.

The rest of the paper has been organized as follows. In Sec.~\ref{El}, the main features of the effective Yukawa sector that induces the flavor violating $Hbs$ coupling are briefly discussed. Sec.~\ref{B} is devoted to derive a constraint on the $Hbs$ vertex from the experimental data on the $B\to X_s\gamma$ decay. In Sec.~\ref{D}, the amplitude for the $b\to s\gamma \gamma$ transition induced by the flavor violating $Hbs$ vertex is introduced and its implications on the $B_s\to \gamma \gamma$ decay are discussed. Finally, in Sec.~\ref{C} the conclusions are presented.

\section{The effective Yukawa sector}
\label{El} As it is shown in Refs.~\cite{T1,T2,T2-1,T3,PV}, it is not necessary to introduce explicitly additional degrees of freedom to generate Higgs-mediated FCNC effects within the SM, but only their virtual effects through an effective Lagrangian that includes $SU_L(2)\times U_Y(1)$-invariant Yukawa-like interactions of dimensions higher than four. An appropriate effective Yukawa sector that generates flavor violating effects in the quark sector is given by~\cite{T1,T2,T2-1,T3,PV}:
\begin{eqnarray}
{\cal L}^Y_{eff}&=&
-Y^d_{ij}(\bar{Q}_i\Phi
d_j)-\frac{\alpha^d_{ij}}{\Lambda^2}(\Phi^\dag \Phi)(\bar{Q}_i\Phi
d_j)+ H.c.\nonumber\\
&&-Y^u_{ij}(\bar{Q}_i\tilde{\Phi}
u_j)-\frac{\alpha^u_{ij}}{\Lambda^2}(\Phi^\dag
\Phi)(\bar{Q}_i\tilde{\Phi} u_j)+ H.c.,
\end{eqnarray}
where $Y_{ij}$, $Q_i$, $\Phi$, $d_i$, and $u_i$
stand for the usual components of the Yukawa matrix, the left-handed quark doublet, the
Higgs doublet, and the right-handed quark singlets of down and up type, respectively.
The $\alpha_{ij}$ numbers are the components of a $3\times 3$
general matrix, which parametrize the details of the underlying
physics, whereas $\Lambda$ is the typical scale of these new
physics effects.

After spontaneous symmetry breaking, this extended Yukawa sector
can be diagonalized in the usual manner via the unitary matrices
$V^{d,u}_L$ and $V^{d,u}_R$, which correlate gauge states to mass
eigenstates. In the unitary gauge, the diagonalized Lagrangian can
be written as follows:
\begin{eqnarray}\label{leff}
{\cal
L}^Y_{eff}&=&-\Big(1+\frac{g}{2m_W}H\Big)\Big(\bar{D}M_dD+\bar{U}M_uU\Big)\nonumber
\\
&&-H\Big(1+\frac{g}{4m_W}H\Big(3+\frac{g}{2m_W}H\Big)\Big)\Big(\bar{D}\Omega^d
P_RD+\bar{U}\Omega^u P_RU+H.c.\Big),
\end{eqnarray}
where the $M_a$ ($a=d,u$) are the diagonal mass matrix and
$\bar{D}=(\bar{d},\bar{s},\bar{b})$ and
$\bar{U}=(\bar{u},\bar{c},\bar{t})$ are vectors in the flavor
space. In addition, $\Omega^a$ are matrices defined in the flavor
space through the relation:
\begin{equation}
\Omega^a=\frac{1}{\sqrt{2}}\Big(\frac{v}{\Lambda}\Big)^2V^a_L\alpha^aV^{a\dag}_R.
\end{equation}
To generate Higgs-mediated FCNC effects at the level of classical
action, it is assumed that neither $Y^{d,u}$ nor
$\alpha^{d,u}$ are diagonalized by the $V^a_{L,R}$ rotation
matrices, which should only diagonalize the sum
$Y^{d,u}+\alpha^{d,u}$. As a consequence, mass and
interactions terms would not be simultaneously diagonalized as it
occurs in the dimension-four theory. In addition, if
$\Omega^{a\dag}\neq \Omega^a$, the Higgs boson couples to fermions
through both scalar and pseudoscalar components, which in turn
could lead to CP violation in some processes. As a consequence,
the flavor violating coupling $H\bar{q}_iq_j$ has the most general renormalizable
structure of scalar and pseudoscalar type given by: $-i(\Omega_{ij}P_R+\Omega^*_{ij}P_L)$.
Notice also that the Lagrangian in Eq.~(\ref{leff}) gives not only couplings of the type $H\bar{q}_iq_j$,
but also couplings of the type $HH\bar{q}_iq_j$, \ldots, etc., being these of no interest for our analysis.

\section{Constraint on $Hbs$ from $B\to X_s\gamma$}
\label{B} The sensitivity of the  $b\to s\gamma$ transition to new physics effects has been studied in diverse approaches beyond the SM, as supersymmetric models~\cite{SUSY}, the two Higgs doublet model~\cite{2HDM}, left-right symmetric models~\cite{LRM}, technicolor models~\cite{Tech}, models with a fourth generation~\cite{FG}, supergravity models~\cite{SG}, effective field theories~\cite{EFT}, the littlest Higgs model~\cite{LHM}, unparticle interactions~\cite{UP}, $331$ models~\cite{331}, and extra dimensions~\cite{ED}. In this section, we calculate the contribution of the flavor violating $Hbs$ coupling to the $b\to s\gamma$ and $b\to sg$ decays and study their implications on the $B\to X_s\gamma$ process. At the leading order (LO) in QCD, the $b\to s$ transition is described via an operator product expansion based on the effective Hamiltonian~\cite{SMA}
\begin{equation}
H_{eff}=-\frac{4G_F}{\sqrt{2}}V^*_{ts}V_{tb}\sum_{i=1}^{8}C_i(\mu)O_i(\mu),
\end{equation}
where the Wilson coefficients $C_i(\mu)$ are evolved from the electroweak scale down to $\mu=m_b$ by the renormalization group equations. The $O_i$ is a set of eight renormalized dimension-six operators. From these, the $O_{1-6}$ represent interactions among four light quarks and are not of interest for our purposes. The remainder $O_7$ and $O_8$ parametrize the electromagnetic dipolar transition and the analogous strong dipolar transition, whose contributions to the $b\to s\gamma$ and $b\to sg$ transitions are dominated by one-loop effects of the $t$ quark and the $W$ gauge boson. The corresponding amplitudes can be written as follows:
\begin{equation}\label{ampsm}
\mathcal{M}_{SM}(b\to s\gamma)=-V_{tb}V^*_{ts}\frac{\alpha^{\frac{3}{2}}}{4\sqrt{\pi}s^2_W m^2_W}C_7(\mu)\bar{s}(p_s)\sigma_{\mu\nu}\epsilon^{*\mu}(q,\lambda)q^\nu(m_s P_L+m_b P_R) b(p_b),
\end{equation}
\begin{equation}
\mathcal{M}_{SM}(b\to sg)=-V_{tb}V^*_{ts}\frac{\sqrt{\alpha_s}\alpha}{4\sqrt{\pi}s^2_W m^2_W}C_8(\mu)\bar{s}(p_s)\sigma_{\mu\nu}\epsilon^{*\mu}_a(q,\lambda)q^\nu T^a(m_s P_L+m_b P_R)b(p_b),
\end{equation}
where $\epsilon^\mu (q,\lambda)$ and $\epsilon^\mu_a(q,\lambda)$ are the polarization vectors of the photon and gluon, respectively. Here, $T^a$ are the generators of the $SU_C(3)$ group, which are normalized as $Tr(T^aT^b)=\delta^{ab}/2$, and $\alpha_s$ is the strong coupling constant.

\begin{figure}
\centering
\includegraphics[width=3.2in]{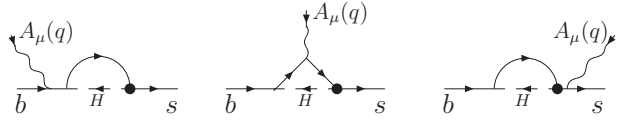}
\caption{\label{FIG1} Diagrams contributing to the $b\to s\gamma$ transition. The $b\to sg$ process occurs via the same type of diagrams.}
\end{figure}

As for new physics effects induced by the flavor violating $Hbs$ vertex, the contribution to the $b\to s\gamma$ and $b\to sg$ transitions is given through the loop diagrams shown in Fig.~\ref{FIG1}. A direct calculation leads to amplitudes of dipolar type, free of ultraviolet divergences, given by
\begin{equation}
\mathcal{M}_{NP}(b\to s\gamma)=-\frac{Q_b\alpha}{16\pi s_W m_W}\mathcal{F}\,\bar{s}(p_s)\sigma_{\mu\nu}\epsilon^{*\mu}(q,\lambda)q^\nu (\Omega_{bs}^*P_L+\Omega_{bs}P_R) b(p_b),
\end{equation}
\begin{equation}
\mathcal{M}_{NP}(b\to sg)=-\frac{\sqrt{\alpha_s\alpha}}{16\pi s_W m_W}\mathcal{F}\,\bar{s}(p_s)\sigma_{\mu\nu}\epsilon_a^{*\mu}(q,\lambda)q^\nu T^a(\Omega_{bs}^*P_L+\Omega_{bs}P_R) b(p_b),
\end{equation}
where $Q_b$ is the electric charge of $b$, $s_W$ is the sine of the weak angle, and $\mathcal{F}$ is the loop function given by
\begin{equation}
\mathcal{F}=\frac{3}{2}-2(B_0(3)-B_0(2))-\frac{m_H^2}{m_b^2}(2B_0(1)-B_0(3)-B_0(4)+2).
\end{equation}
Here, $B_0(i)$ are Passarino-Veltman~\cite{PV} scalar functions given by:  $B_0(1)=B_0(0, m_H^2, m_H^2)$, $B_0(2)=B_0(0, m_b^2, m_b^2)$, $B_0(3)=B_0(0, m_H^2, m_b^2)$, and $B_0(4)=B_0(m_b^2, m_H^2, m_b^2)$. Let us mention that we are tacitly assuming that $\Omega_{bs}$ is purely imaginary; otherwise the $CP$-violation condition is not satisfied as it is required for the $B_s$ meson.

In the context of the effective theory that we are considering, the total theoretical contribution to the $b-s$ transition is given by the sum of the SM contribution and the new physics effect induced by the $Hbs$ vertex:
\begin{equation}
{\cal M}_T={\cal M}_{SM}+{\cal M}_{NP}.
\end{equation}

\begin{figure}
\centering
\includegraphics[width=4.0in]{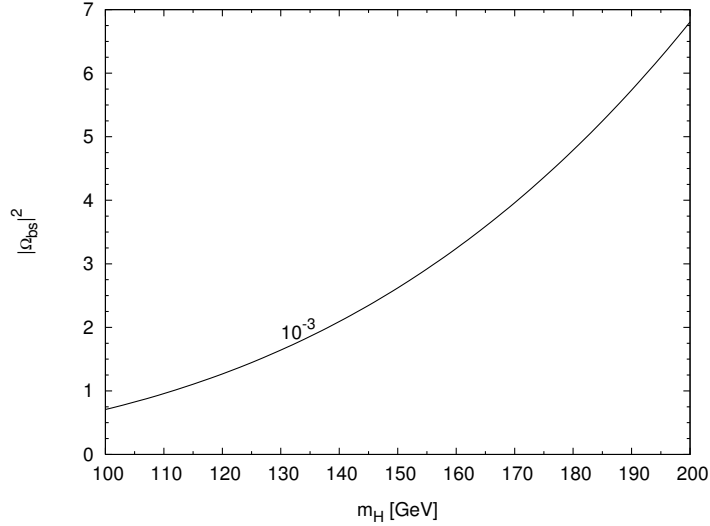}
\caption{\label{FIG2} $|\Omega_{bs}|^2$ as a function of the Higgs mass.}
\end{figure}

Our main objective in this section is to get a bound for the $\Omega_{bs}$ parameter. We will follow closely the analysis given in Ref.~\cite{UP}. The discrepancy between the theoretical prediction within the SM and the experimental measurement can be quantified via the following ratio:
\begin{equation}\label{disc}
R_{EXP-SM}\equiv \frac{\Gamma_{EXP}-\Gamma_{SM}}{\Gamma_{SM}}=\frac{Br_{EXP}(B\to X_s\gamma)}{Br_{SM}(B\to X_s\gamma)}-1,
\end{equation}
 where $\Gamma_{EXP}$ is the experimental decay width of the $B\to X_s\gamma$ transition and  $\Gamma_{SM}$ is the corresponding theoretical prediction of the SM. In addition, $Br_{EXP}$ and $Br_{SM}$ are the respective branching ratios. Using the experimental and theoretical values presented at the beginning of the introduction, it is found that the discrepancy between theory and experiment is given by $R_{EXP-SM}=0_\cdot117\pm 0_\cdot113$. To constraint the $Hbs$ vertex, we will assume that the total theoretical prediction, \textit{i.e.}, the SM prediction plus the $Hbs$ contribution, coincides with the experimental value. Thus, we define the ratio
\begin{equation}\label{discota}
R_{TOT-SM}\equiv \frac{\Gamma_{SM+NP}-\Gamma_{SM}}{\Gamma_{SM}}=\frac{Br_{SM+NP}}{Br_{SM}}-1,
\end{equation}
which quantifies the theoretical discrepancy between the effective theory prediction (SM plus new physics effects) and the SM prediction. We now demand that $R_{TOT-SM}\approx R_{EXP-SM}$, which allows us to obtain a bound for the $\Omega_{bs}$ parameter. Before to do this, some additional considerations are needed. Working out at LO, which is sufficient for our purposes, the SM contribution can be written as follows:
\begin{equation}
\mathcal{M}_{SM}(b\to s\gamma)=\bar{s}(p_s)\sigma_{\mu\nu}\epsilon^{*\mu}(q,\lambda)q^\nu(A^L_{SM} P_L+A^R_{SM} P_R) b(p_b),
\end{equation}
where the $A^{L, R}_{SM}$ form factors are given by
\begin{align}\label{Asm}
A^{L}_{SM}&=-V_{tb}V^*_{ts}\frac{\alpha^{\frac{3}{2}}}{4\sqrt{\pi}s^2_W m^2_W}C^{eff}_7(m_b)\,m_s,\nonumber\\
A^{R}_{SM}&=-V_{tb}V^*_{ts}\frac{\alpha^{\frac{3}{2}}}{4\sqrt{\pi}s^2_W m^2_W}C^{eff}_7(m_b)\,m_b,
\end{align}
with an effective Wilson coefficient $C^{eff}_7(m_b)=0_\cdot689C_7(m_W)+0_\cdot087C_8(m_W)$~\cite{SMA}, which already contains the QCD contribution at the $m_b$ scale. In a similar way, the corresponding new physics contribution can be written as follows:
\begin{equation}
\mathcal{M}_{NP}(b\to s\gamma)=\bar{s}(p_s)\sigma_{\mu\nu}\epsilon^{*\mu}(q,\lambda)q^\nu (A^L_{NP} P_L+A^R_{NP} P_R) b(p_b),
\end{equation}
where
\begin{align}\label{Anp}
A^{L}_{NP}&=-\frac{Q_b\alpha}{16\pi s_W m_W}\Omega_{bs}^\ast\mathcal{F}\,\bigg(0_\cdot689+\frac{0_\cdot087}{Q_b}\bigg),\nonumber\\
A^{R}_{NP}&=-\frac{Q_b\alpha}{16\pi s_W m_W}\Omega_{bs}\mathcal{F}\,\bigg(0_\cdot689+\frac{0_\cdot087}{Q_b}\bigg),
\end{align}

From expression (\ref{discota}) and the assumption $R_{TOT-SM}\approx R_{EXP-SM}$, one obtains
\begin{equation}
R_{EXP-SM}=\frac{|A^L_{SM}+A^L_{NP}|^2+|A^R_{SM}+A^R_{NP}|^2}{|A^L_{SM}|^2+|A^R_{SM}|^2}-1.
\end{equation}
The problem of finding a bound for the $\Omega_{bs}$ parameter reduces now to solve a quadratic equation, which has two solutions\footnote{In the numerical evaluation, the central value for $R_{EXP-SM}$ was used.}. The physical solution corresponds to that for which the allowed values for $\Omega_{bs}$ satisfy the $|A_{SM}|^2>|A_{NP}|^2$ condition, which is a reasonable requirement. In Fig.~\ref{FIG2}, the behavior of $|\Omega_{bs}|^2$ as a function of the Higgs mass in the range $100$ GeV $<m_H< 200$ GeV is shown. From this figure, it can be appreciated that $|\Omega_{bs}|^2<(0.7-6.8)\times 10^{-3}$ for a Higgs mass in the range $115$ GeV$<m_H<200$ GeV. On the other hand, we would like to mention that in Ref.~\cite{T2-1} experimental results from $D^0-\bar{D}^0$ mixing were used to get a constraint on the $Htc$ coupling, where the parameter $\Omega_{tc}$ was bounded to be $\Omega_{tc}^2<10^{-2}$, which represents the strength of the $Htc$ interaction.

\section{The $B_s\to \gamma \gamma$ decay}
\label{D}

\begin{figure}
\centering
\includegraphics[width=3.6in]{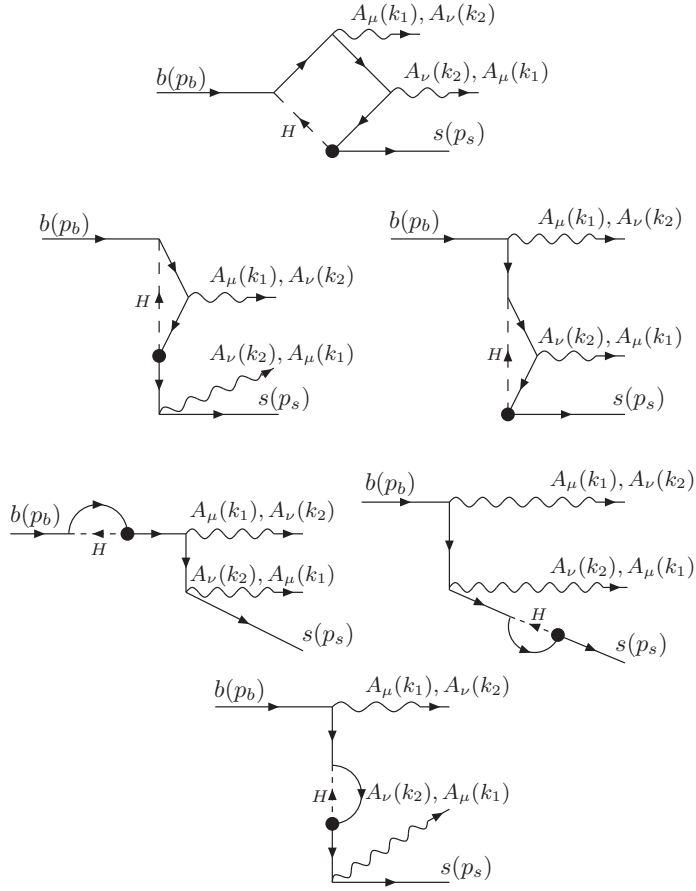}
\caption{\label{diabox-red}Contribution of the box and reducible diagrams to the $b\to s\gamma\gamma$ decay.}
\end{figure}

\begin{figure}
\centering
\includegraphics[width=4.0in]{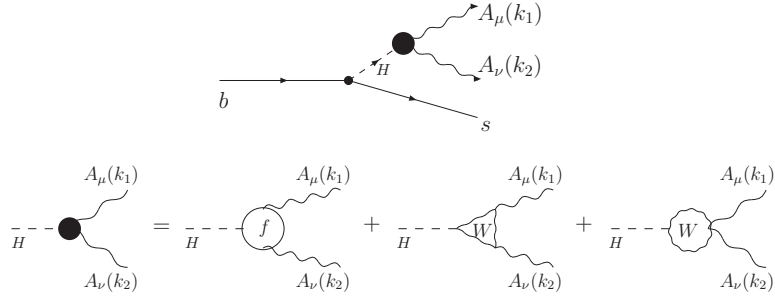}
\caption{\label{diauni}Contribution of the SM one-loop induced $H^\ast\gamma\gamma$ vertex to the $b\to s\gamma\gamma$ decay.}
\end{figure}

As mentioned above, the $Hbs$ effective vertex induces the flavor violating process $b\to s\gamma\gamma$ at the one-loop level (with kinematics defined in Fig.~\ref{diabox-red}). The contribution to $b\to s\gamma\gamma$ occurs through two set of Feynman diagrams, each given a finite and gauge invariant contribution~\cite{T3}. The first set of diagrams (see Fig.~\ref{diabox-red}) includes box diagrams, reducible diagrams characterized by the one-loop $bs\gamma$ coupling, and reducible diagrams composed by the one-loop $b-s$ bilinear coupling. Henceforth we will refer to this set of graphs as box-reducible diagrams. The second set of diagrams is characterized by the SM one-loop $H^\ast \gamma\gamma$ coupling, where $H^\ast$ represents a virtual Higgs boson (see Fig.~\ref{diauni}). These type of graphs will be named Higgs-reducible diagrams. We find that the dominant Higgs mediated flavor violating effect is given by the contribution of the Feynman diagrams shown in Fig.~\ref{diauni}. The arising contributions from box-reducible graphs are marginal due to the bottom quark charge, which induces an extra factor equal to 1/9 at the one-loop level amplitude.

In order to make predictions, we will use our previous result for the flavor violating parameter $\Omega_{bs}(m_H)$ as a function of the Higgs mass. We find that the amplitude for the $b\to s\gamma\gamma$ decay can be written as~\cite{T3}:

\begin{equation}\label{AmpBsr}
\mathcal{M}^{\mu\nu}=\frac{\alpha\,g}{8 \pi m_W}\,F_0\,\bar{u}_s(p_s)(\Omega_{bs}P_R+\Omega^\ast_{bs}P_L)\frac{k_2^\mu k_1^\nu-k_1\cdot k_2 g^{\mu\nu}}{2k_1\cdot k_2-m_H^2+im_H\Gamma_H}u_b(p_b),
\end{equation}
with

\begin{align}
F_0=\frac{8\,m_W^2}{2\,k_1\cdot k_2}\left(3 + \frac{2\,k_1\cdot k_2}{2\,m_W^2} +
6\,m_W^2\left(1 - \frac{2\,k_1\cdot k_2}{2\,m_W^2}\right)\,C_0(1)\right)-Q_t^2\,{N_{c_{t}}}\,\frac{8\,m_t^2}{2\,k_1\cdot k_2}
\left(2 + (4\,m_t^2 - 2\,k_1\cdot k_2)\,C_0(2)\right),
\end{align}
where $C_0(1)=C_0(0, 0, 2\,k_1\cdot k_2, m_W^2, m_W^2, m_W^2)$ and $C_0(2)=C_0(0, 0, 2\,k_1\cdot k_2, m_t^2, m_t^2, m_t^2)$ are the Passarino-Veltman scalar functions, $m_t$ is the top quark mass, $Q_t$ is the top quark charge, and $N_{c_{t}}=3$ is the color factor.

According to the static quark approximation~\cite{SQA, BRPSUSY}, we can compute the decay width $\Gamma(B_s\to\gamma\gamma)$ starting from $\Gamma(b\to s\gamma\gamma)$, where it is assumed that the three-momenta of the $b$ and $s$ quarks vanish in the rest frame of the $B_s$ meson. In this approximation, the $B_s$ meson decays into two photons emitted with energies $m^2_{B_s}/2$ and the product $k_1\cdot k_2=m_{B_s}^2/2$, where $m_{B_s}=m_b+m_s$~\footnote{As in Refs.~\cite{SQA, BRPSUSY}, we will use the constituent mass for the strange quark $m_s=m_K=0.497$ GeV.} is the $B_s$ meson mass.

In order to get the amplitude for $B_s\to\gamma\gamma$ we resort to the following matrix elements~\cite{SQA, BRPSUSY}:

\begin{align}
\langle0|\bar u_s\gamma^5 u_b|B_s\rangle&=if_{B_s}m_{B_s},\nonumber\\
\langle0|\bar u_s\gamma^\mu\gamma^5 u_b|B_s\rangle&=if_{B_s}P^\mu,
\end{align}
where $P=p_b-p_s$ is the $B_s$-meson four-momentum and $f_{B_s}$ is the $B_s$-meson decay constant. By using the above matrix elements the amplitude for $B_s\to \gamma\gamma$ can be written as follows:

\begin{equation}
\mathcal{M}^{\mu\nu}(B_s\to \gamma\gamma)=f_{B_S}B_{NP}(k_1^\nu\cdot k_2^\mu-\frac{m_{B_s}^2}{2}g^{\mu\nu}),
\end{equation}
where $B_{NP}$ is the Higgs mediated flavor violating form factor defined as

\begin{equation}
B_{NP}=\frac{\alpha^{\frac{3}{2}} \Omega_{bs}}{4\pi^{\frac{1}{2}} s_W}\frac{m_{B_s}}{m_W m_H^2}\,F_0.
\end{equation}
This implies that the decay width for $B_s\to \gamma\gamma$ process arising from the new physics effects encoding in $B_{NP}$ has the following form

\begin{equation}
\Gamma(B_s\to \gamma\gamma)=f_{B_{s}}^2\frac{m^3_{B_{s}}}{16\pi}|B_{NP}|^2.
\end{equation}
Finally, we can compute the corresponding branching ratio for $B_s\to\gamma\gamma$ decay by means of

\begin{equation}
Br(B_s\to \gamma\gamma)= \frac{\Gamma(B_s\to \gamma\gamma)}{\Gamma_{T}(B_s)},
\end{equation}
where $\Gamma_{T}(B_s)$ is the total $B_s$-meson width decay determined by its lifetime $\tau(B_s)=1_\cdot43$ ps~\cite{PDG}.

\begin{figure}
\centering
\includegraphics[width=4.0in]{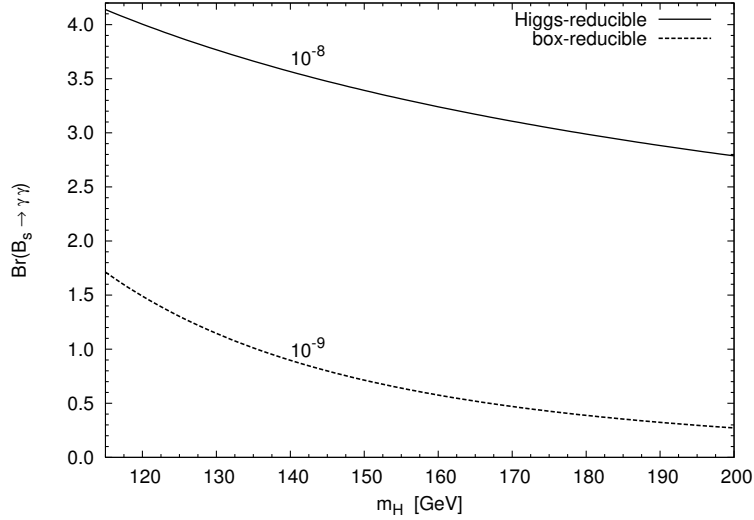}
\caption{\label{brBs2body}The branching ratio of the $B_s\to\gamma\gamma$ decay for the Higgs-reducible contribution (solid) and box-reducible contribution (dashed) as a function of the Higgs mass.}
\end{figure}

Hereafter, we shall present our results for $Br(B_s\to\gamma\gamma)$ as a function of the Higgs mass. The results were obtained using the values $m_{B_s}=5.37$ GeV, $m_{K}=0.497$ GeV, and $f_{B_s}=0.24$ GeV. We show in Fig.~\ref{brBs2body} the branching ratio for the $B_s\to\gamma\gamma$ process, where it has displayed separately the contributions coming from Higgs-reducible and box-reducible diagrams. From this figure, it can be appreciated that the contribution induced by the Higgs-reducible graphs is approximately 2 orders of magnitude larger than those generated by the box-reducible graphs in the range of a Higgs mass of 115 GeV$<m_H<$ 200 GeV. Moreover, we can see that $Br(B_s\to\gamma\gamma)$ ranges from $4.14\times 10^{-8}$ to $2.79\times 10^{-8}$ for a Higgs mass in the same range. We note that the behavior of both curves in Fig.~\ref{brBs2body} is such that, essentially, the difference in order of magnitude is maintained along the Higgs mass range analyzed.

From the discussion presented in the introduction, we can appreciate that our prediction for the Higgs mediated flavor violating $Br(B_s\to\gamma\gamma)$ process is almost 2 orders of magnitude more stringent that the current experimental limit and 1 order of magnitude lower than those derived from SM and its extensions.

\section{Conclusions}
\label{C}

Effective theories beyond the SM  can encode new physics effects such as flavor violating transitions which is of current interest. In this work we have used dimension-six operators in the Yukawa sector to study these transitions mediated by a SM-like Higgs boson. In particular we have studied the resulting $Hbs$ coupling and estimated its strength  from the branching ratio for the $B\to X_s\gamma$ process, specifically, the effective parameter $\Omega_{bs}$ was bounded by using the discrepancy between the respective theoretical and experimental  central values of the branching ratios. This constraint was used to bound the Higgs mediated flavor violating $B_s\to\gamma\gamma$ decay and we found that its branching ratio is less than $10^{-8}$ in the Higgs mass interval ranging from 115 GeV to 200 GeV. Our results on the branching ratio in question are 2 orders of magnitude more stringent than the current experimental bound imposed by the Belle Collaboration. As expected we found that the box-reducible diagrams, which take into account the short-distance effects contribute marginally to the $B_s\to\gamma\gamma$ process, being this dominated by the contribution of the Higgs-reducible diagrams. Finally we would like to stress that our bound for this process is model-independent, with two free parameters, namely the effective coupling strength $\Omega_{bs}$ which is fixed from experimental results for the branching ratio of the $B\to X_s\gamma$ process, and the mass of the Higgs boson.

\acknowledgments{We would like to thank J. L. D\'\i az-Cruz for his comments. We acknowledge financial support from CONACYT and
SNI (M\' exico).}


\begin{thebibliography}{99}
%
\bibitem{Review} For a recent review, see T. Aushev \textit{et al.}, arXiv:1002.5012 [hep-ex].
%
\bibitem{Hewett} J. L. Hewett, hep-ph/9406302.
%
\bibitem{HFAG} E. Barberio \emph{et al.} [Heavy Flavor Averaging Group], arXiv:0808.1297 [hep-ex].
%
\bibitem{SMP} M. Misiak \emph{et al.}, Phys. Rev. Lett. 98, 022002 (2007); M. Misiak and M. Steinhauser, Nucl. Phys. \textbf{B764}, 62 (2007).
%
\bibitem{EL} W. Buchmuller and D. Wyler, Nucl. Phys. \textbf{B268}, 621 (1986). See also, J. Wudka, Int. J. Mod. Phys. \textbf{A9}, 2301 (1994).
%
\bibitem{T1} J. L. D\'\i az-Cruz and J. J. Toscano, Phys. Rev. \textbf{D62}, 116005 (2000); A. Cordero-Cid, M. A. P\'erez, G. Tavares-Velasco, and J. J. Toscano, Phys. Rev. \textbf{D70}, 074003 (2004).
 %
\bibitem{T2} J. I. Aranda, F. Ram\'\i rez-Zavaleta, J. J. Toscano, and  E. S. Tututi, Phys. Rev. \textbf{D78}, 017302 (2008).
%
\bibitem{T2-1} J. I. Aranda, A. Cordero-Cid, F. Ram\'\i rez-Zavaleta, J. J. Toscano, and E. S. Tututi, Phys. Rev. \textbf{D81}, 077701 (2010); A. Fern\'andez, C. Pagliarone, F. Ram\'\i rez-Zavaleta, and J. J. Toscano, J. Phys. G, in press, arXiv:0911.4995.
 %
\bibitem{T3} J. I. Aranda, A. Flores-Tlalpa, F. Ram\'\i rez-Zavaleta, F. J. Tlachino, J. J. Toscano, and E. S. Tututi, Phys. Rev. \textbf{D79}, 093009 (2009).
%
\bibitem{PV} F. del Aguila, M. P\' erez-Victoria, and J. Santiago, Phys. Lett. \textbf{B492}, 98 (2000).
%
\bibitem{THDM-III} D. Atwood, L. Reina, and A. Soni, Phys. Rev. \textbf{D55}, 3156 (1997); J. F. Gunion and H. E. Haber, Phys. Rev. \textbf{D67}, 075019 (2003); \textit{ibid.} \textbf{72}, 095002 (2005); S. Davidson and H. E. Haber, Phys. Rev. \textbf{D72}, 035004 (2005); M. Aoki, S. Kanemura, K. Tsumura, and K. Yagyu, Phys. Rev. \textbf{D80}, 015017 (2009).
%
\bibitem{Familons} J. L. Feng, T. Moroi, H. Murayama, and E. Schnapka, Phys. Rev. \textbf{D57}, 5875 (1998).
%
\bibitem{AFS} I. Dorsner and S. M. Barr, Phys. Rev. \textbf{D65}, 095004 (2002).
%
\bibitem{Yao} Guey-Lin Lin, Jiang Liu, and York-Peng Yao, Phys. Rev. Lett. \textbf{64}, 1498 (1990); Phys. Rev. \textbf{D42}, 2314 (1990).
%
\bibitem{SQA} S. Herrlich and J. Kalinowski, Nucl. Phys. \textbf{B381}, 501 (1992).
%
\bibitem{NOQCD} Chia-Hung V. Chang, Guey-Lin Lin, and York-Peng Yao,  Phys. Lett. \textbf{B415}, 395 (1997);  L. Reina, G. Ricciardi, and A. Soni, Phys. Rev. \textbf{D56}, 5805 (1997).
%
\bibitem{LD2} S. W. Bosch, G. Buchalla, JHEP 0208, 054 (2002).
%
\bibitem{LD1}  G. Hiller and E. O. Iltan, Phys. Lett. \textbf{B409}, 425 (1997).
%
\bibitem{LD3} D. Choudhury and J. Ellis, Phys. Lett. \textbf{B433}, 102 (1998).
%
\bibitem{SBSUSY} S. Bertolini and J. Matias, Phys. Rev. \textbf{D57}, 4197 (1998).
%
\bibitem{THDM} T. M. Aliev, G. Hiller, and E. O. Iltan, Nucl. Phys. \textbf{B515}, 321 (1998).
%
\bibitem{FG} W.~j.~Huo, C.~D.~Lu and Z.~j.~Xiao, hep-ph/0302177.
%
\bibitem{BRPSUSY} A. Gemintern, S. Bar-Shalom, and G. Eilam, Phys. Rev. \textbf{D70}, 035008 (2004).
%
\bibitem{BelleExp} J. Wicht \textit{et al.} (Belle Collaboration), Phys. Rev. Lett. \textbf{100}, 121801 (2008).
%
\bibitem{SUSY}V. Barger, M. S. Berger, and R. J. N. Phillips, Phys. Rev. Lett. \textbf{70}, 1368 (1993); J. L. Hewett, Phys. Rev. Lett. \textbf{70}, 1045 (1993); M. A. D\'\i az, Phys. Lett. \textbf{B304}, 278 (1993); R. Barbieri and G. F. Giudice, Phys. Lett. \textbf{B309}, 86 (1993); Y. Okada, Phys. Lett. \textbf{B315}, 119 (1993); R. Garisto and J. N. Ng, Phys. Lett. \textbf{B315}, 372 (1993); F. M. Borzumati, Z. Phys. \textbf{C63}, 291 (1994); S. Bertolini and F. Vissani, Z. Phys. \textbf{C67}, 513 (1995); H. Anlauf, Nucl. Phys. \textbf{B430}, 245 (1994); V. Barger, M. S. Berger, P. Ohmann, and R. J. N. Phillips, Phys. Rev. \textbf{D51}, 2438 (1995); P. Nath and R. L. Arnowitt, Phys. Rev. Lett. \textbf{74}, 4592 (1995); F. M. Borzumati, M. Olechowski, and S. Pokorski, Phys. Lett. \textbf{B349}, 311 (1995); N. G. Deshpande, B. Dutta, and S. Oh, Phys. Rev. \textbf{D56}, 519 (1997); T. Bla$\check{\mathrm{z}}$ek and S. Raby, Phys. Rev. \textbf{D59}, 095002 (1999); H. Baer, M. Brhlik, D. Casta\~no, and X. Tata, Phys. Rev. \textbf{D58}, 015007 (1998); W. de Boer, H.-J. Grimm, A. V. Gladyshev, and D. I. Kazakov, Phys. Lett. \textbf{B438}, 281 (1998); M. A. D\'\i az, E. Torrente-Lujan, and J. W. F. Valle, Nucl. Phys. \textbf{B551}, 78 (1999); Chun-Khiang Chua, Xiao-Gang He, and Wei-Shu Hou, Phys. Rev. \textbf{D60}, 014003 (1999); Y. G. Kim, P. Ko, and J. S. Lee, Nucl. Phys. \textbf{B544}, 64 (1999); E. Ma and M. Raidal, Phys. Lett. \textbf{B491}, 297 (2000);  T. Besmer and A. Steffen, Phys. Rev. \textbf{D63}, 055007 (2001); E. Gabrielli, S. Khalil, and E. Torrente-Lujan, Nucl. Phys. \textbf{B594}, 3 (2001); D. Chakraverty and D. Choudhury, Phys. Rev. \textbf{D63}, 075009 (2001); M. Carena, D. Garcia, U. Nierste, and C. E. M. Wagner, Phys. Lett. \textbf{B499}, 141 (2001); M. B. Causse and J. Orloff, Eur. Phys. J. \textbf{C23}, 749 (2002); S. Komine and M. Yamaguchi, Phys. Rev. \textbf{D65}, 075013 (2002); L. Everett, G. L. Kane, S. Rigolin, Lian-Tao Wang, and T. T. Wang, JHEP \textbf{0201}, 022 (2002); Ken-ichi Okumura and L. Roszkowski, Phys. Rev. Lett. \textbf{92}, 161801 (2004); M. E. G\'omez, T. Ibrahim, P. Nath, and S. Skadhauge, Phys. Rev. \textbf{D74}, 015015 (2006); L. Roszkowski, R. Ruiz de Austri, and R. Trotta, JHEP \textbf{0707}, 075 (2007); F. Mahmoudi, JHEP \textbf{0712}, 026 (2007); K. A. Olive and L. Velasco-Sevilla, JHEP \textbf{0805}, 052 (2008);  Y. Yamada, Phys. Rev. \textbf{D77}, 014025 (2008); B. Dudley and C. Kolda, Phys. Rev. \textbf{D79}, 015011 (2009); N. Chen, D. Feldman, Z. Liu, and P. Nath, Phys. Lett. \textbf{B685}, 174 (2010).
%
\bibitem{2HDM} T. G. Rizzo, Phys. Rev. \textbf{D38}, 820 (1988); G. T. Park, Phys. Rev. \textbf{D50}, 599 (1994); Mod. Phys. Lett. \textbf{A9}, 321 (1994); L. Wolfenstein and Y. L. Wu, Phys. Rev. Lett. \textbf{73}, 2809 (1994); Cai-Dian L$\mathrm{\ddot{u}}$, Nucl. Phys. \textbf{B441}, 33 (1995); T. M. Aliev and E. O. Iltan, J. Phys. \textbf{G25}, 989 (1999); D. Bowser-Chao, K. Cheung, and Wai-Yee Keung, Phys. Rev. \textbf{D59}, 115006 (1999).
%
\bibitem{LRM} P. Cho and M. Misiak, Phys. Rev. \textbf{D49}, 5894 (1994); K. S. Babu, K. Fujikawa, and A. Yamada, Phys. Lett. \textbf{B333}, 196 (1994); T. G. Rizzo, Phys. Rev. \textbf{D50}, 3303 (1994); G. Bhattacharyya and A. Raychaudhuri, Phys. Lett. \textbf{B357}, 119 (1995); H. M. Asatrian and A. N. Ioannissian, Phys. Rev. \textbf{D54}, 5642 (1996); C. S. Kim and Y. G. Kim, Phys. Rev. \textbf{D61}, 054008 (2000); M. Frank and S. Nie, Phys. Rev. \textbf{D65}, 114006 (2002).
%
\bibitem{Tech} L. Randall and R. Sundrum, Phys. Lett. \textbf{B312}, 148 (1993); C. D. Carone, E. H. Simmons, and Y. Su, Phys. Lett. \textbf{B344}, 287 (1995); B. Balaji, Phys. Rev. \textbf{D53}, 1699 (1996); G. Lu, Y. Cao, Z. Xiong, and C. Yue, Z Phys. \textbf{C74}, 355 (1997).
%
\bibitem{FG} Wei-Shu Hou, A. Soni, and H. Steger, Phys. Lett. \textbf{B192}, 441 (1987); L. T. Handoko and T. Morozumi, Mod. Phys. Lett. \textbf{A10}, 309 (1995), \emph{ibid.} \textbf{A10}, 1733 (1995); T. M. Aliev, D. A. Demir, and N. K. Pak, Phys. Lett. \textbf{B389}, 83 (1996).
%
\bibitem{SG} A. Masiero and G. Ridolfi, Phys. Lett. \textbf{B212}, 171 (1988), \emph{Addendum-ibid.} \textbf{B213}, 562 (1988); J. L. Lopez, D. V. Nanopoulos, and G. T. Park, Phys. Rev. \textbf{D48}, R974 (1993); M. A. D\'\i az, Phys. Lett. \textbf{B322}, 207 (1994); P. Nath and R. Arnowitt, Phys. Lett. \textbf{B336}, 395 (1994); J. Wu, R. Arnowitt, and P. Nath, Phys. Rev. \textbf{D51}, 1371 (1995); J. L. Lopez, D. V. Nanopoulos, X. Wang, and A. Zichichi, Phys. Rev. \textbf{D51}, 147 (1995).
%
\bibitem{EFT} Swee-Ping Chia, Phys. Lett. \textbf{B240}, 465 (1990);  K. A. Peterson, Phys. Lett. \textbf{B282}, 207 (1992); J. L. Hewett and T. G. Rizzo, Phys. Rev. \textbf{D49}, 319 (1994); T. G. Rizzo, Phys. Lett. \textbf{B315}, 471 (1993); Xiao-Gang He and B. McKellar, Phys. Lett. \textbf{B320}, 165 (1994); R. Mart\'\i nez, M. A. P\' erez, and J. J. Toscano, Phys. Lett. \textbf{B340}, 91 (1994); R. Mart\'\i nez and J.-Alexis Rodr\'\i guez, Phys. Rev. \textbf{D55}, 3212 (1997).
%
\bibitem{LHM} W. Huo and S. Zhu, Phys. Rev. \textbf{D68}, 097301 (2003).
%
\bibitem{UP}  Xiao-Gang He and Lu-Hsing Tsai, JHEP \textbf{0806}, 074 (2008).
%
\bibitem{331} J. Agrawal, P. H. Frampton, and J. T. Liu, Int. J. Mod. Phys. \textbf{A11}, 2263 (1996).
%
\bibitem{ED} K. Agashe, N. G. Deshpande, and G.-H. Wu, Phys. Lett. \textbf{B514}, 309 (2001).
%
\bibitem{SMA} B. Grinstein, R. Springer, and M. B. Wise, Nucl. Phys. \textbf{B339}, 269 (1990); A. J. Buras \emph{et al.}, Nucl. Phys. \textbf{B424}, 374 (1994).
%
\bibitem{PDG} C. Amsler \textit{et al.}, Phys. Lett. \textbf{B667}, 1 (2008).
%
\end{thebibliography}
\end{document}